\newcommand{\jpsi}{\ensuremath{J\kern-0.05em /\kern-0.05em\psi} }
\newcommand{\sqrts}[1]{\ensuremath{\sqrt{s} = #1\,\mathrm{TeV}}}
\newcommand{\pT}{\ensuremath{p_\mathrm{T}} }

\documentclass[3p,times,procedia]{elsarticle}
\usepackage{nupha_ecrc}

\usepackage{ mathrsfs }

\volume{00}

\firstpage{1}

\journalname{Nuclear Physics A}

\runauth{S. G. Weber on behalf of the ALICE Collaboration}

\jid{nupha}

\jnltitlelogo{Nuclear Physics A}

\usepackage{amssymb}

\usepackage[figuresright]{rotating}

\begin{document}

\begin{frontmatter}



\dochead{XXVIth International Conference on Ultrarelativistic Nucleus-Nucleus Collisions\\ (Quark Matter 2017)}

\title{Measurement of \jpsi production as a function of event multiplicity in pp collisions at \sqrts{13} with ALICE}


\author{S. G. Weber on behalf of the ALICE Collaboration}

\address{Research Division, GSI Helmholtzzentrum f\"ur Schwerionenforschung, Darmstadt, Germany}

\begin{abstract}
The availability at the LHC of the largest collision energy in pp collisions allows a significant advance in the measurement of \jpsi production as function of event multiplicity. The interesting relative increase observed with data at the LHC at \sqrts{7} and at RHIC at \sqrts{0.2} is studied now at unprecedented multiplicities at \sqrts{13}. The measurement at mid-rapidity in the dielectron channel with ALICE is presented and discussed in comparison to predictions from available theoretical models and to data at lower energies.

\end{abstract}


\end{frontmatter}


\section{Introduction}

Hadronic charmonium production at collider energies is a complex and not yet fully understood process, involving hard-scale processes, i.e. the initial heavy quark pair production, which can be described by means of perturbative Quantum ChromoDynamics (pQCD), as well as soft scale processes, i.e. the subsequent binding into a charmonium state. A comprehensive description of the transverse momentum and rapidity dependent production down to zero transverse momentum \pT can be obtained with the Non-Relativistic Quantum ChromoDynamics (NRQCD) formalism combined with a Color Glass Condensate (CGC) description of the incoming protons \cite{nrqcd_cgc_13tev}.

The event multiplicity dependent production of charmonium, and also of charm quarks in general, is a fairly new topic of interest, and is believed to be suited to give new insights on processes on the parton level and on the interplay between the hard and soft mechanisms in particle production.

ALICE has performed multiplicity dependent measurements in pp collisions at \sqrts{7} of inclusive \jpsi production at mid- and forward rapidity \cite{jpsi7tev}, and prompt \jpsi, non-prompt \jpsi and D-mesons production at mid-rapidity \cite{dmesons}. The general observation is an increase of open and hidden charm production with event multiplicity. For the \jpsi production, multiplicities of about 4 times the mean values could be reached. The results are consistent with a linear, or stronger than linear increase. For the D-meson production, relative multiplicities of about 6 could be reached, with a stronger than linear increase at the highest multiplicities. Similar observations have been made by CMS for $\Upsilon(nS)$ particles at mid-rapidity. They find a linear increase with event activity, when measuring it at forward rapidity, and a stronger than linear increase when measuring it at mid-rapidity \cite{upsilon}.

Different theoretical models attribute the observed behavior to different underlying processes. In the PYTHIA 8 event generator \cite{pythia}, multi-parton interactions (MPI) are an important factor in charm quark production. Indeed, from MPI alone a stronger than linear scaling is expected for open charm production, as was demonstrated in \cite{dmesons}. Taking into account all sources of heavy quark production however, a more or less linear increase is predicted. In the EPOS 3 event generator \cite{epos3}, initial conditions are generated according to the ”Parton based Gribov-Regge” formalism \cite{epos} followed by a hydrodynamical evolution of the system. Sources of particle production in this framework are parton ladders, each composed of a pQCD hard process with initial and final state radiation. This model predicts a stronger than linear increase with multiplicity \cite{dmesons}.

In the percolation model \cite{percolation}, spatially extended color strings are the sources of particle production in high energy hadronic collisions. In a high density environment they overlap, decreasing their effective number. Since the transverse size of the strings is determined by their transverse mass, light sources are affected in a stronger way than the heavy ones. This results in a linear increase of heavy particle production at low multiplicities, gradually changing to a quadratic one at high multiplicities.

The model by Kopeliovich et al. \cite{kopeliovich} assumes that high multiplicities are the results of contributions from higher Fock states in the incoming protons, leading to a large number of gluons participating in a collision. An increase of \jpsi production with the event multiplicity is expected.

\section{Experimental setup and analysis methods}

\begin{figure}[t]
\begin{minipage}{0.6\textwidth}
\includegraphics[width=\textwidth]{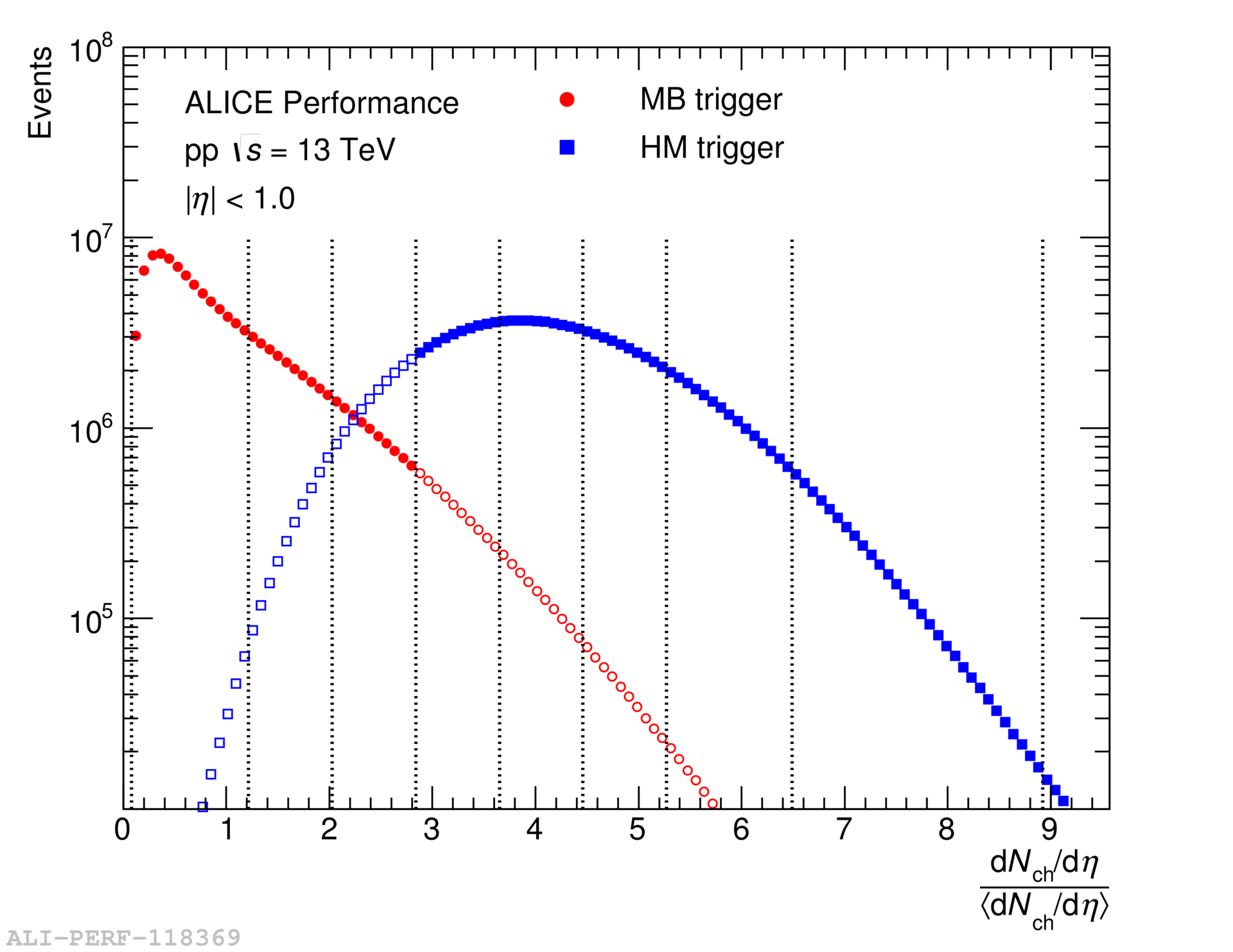}
\caption{Multiplicity distribution of events using a minimum bias trigger, or a high multiplicity trigger, respectively. The filled symbols indicate the events used for this analysis, the dotted black lines show the limits of the bins used for the signal extraction. \label{mult}}
\end{minipage}
\begin{minipage}{0.01\textwidth}
\hspace{12pc}
\end{minipage}
\begin{minipage}{0.39\textwidth}
\includegraphics[width=\textwidth]{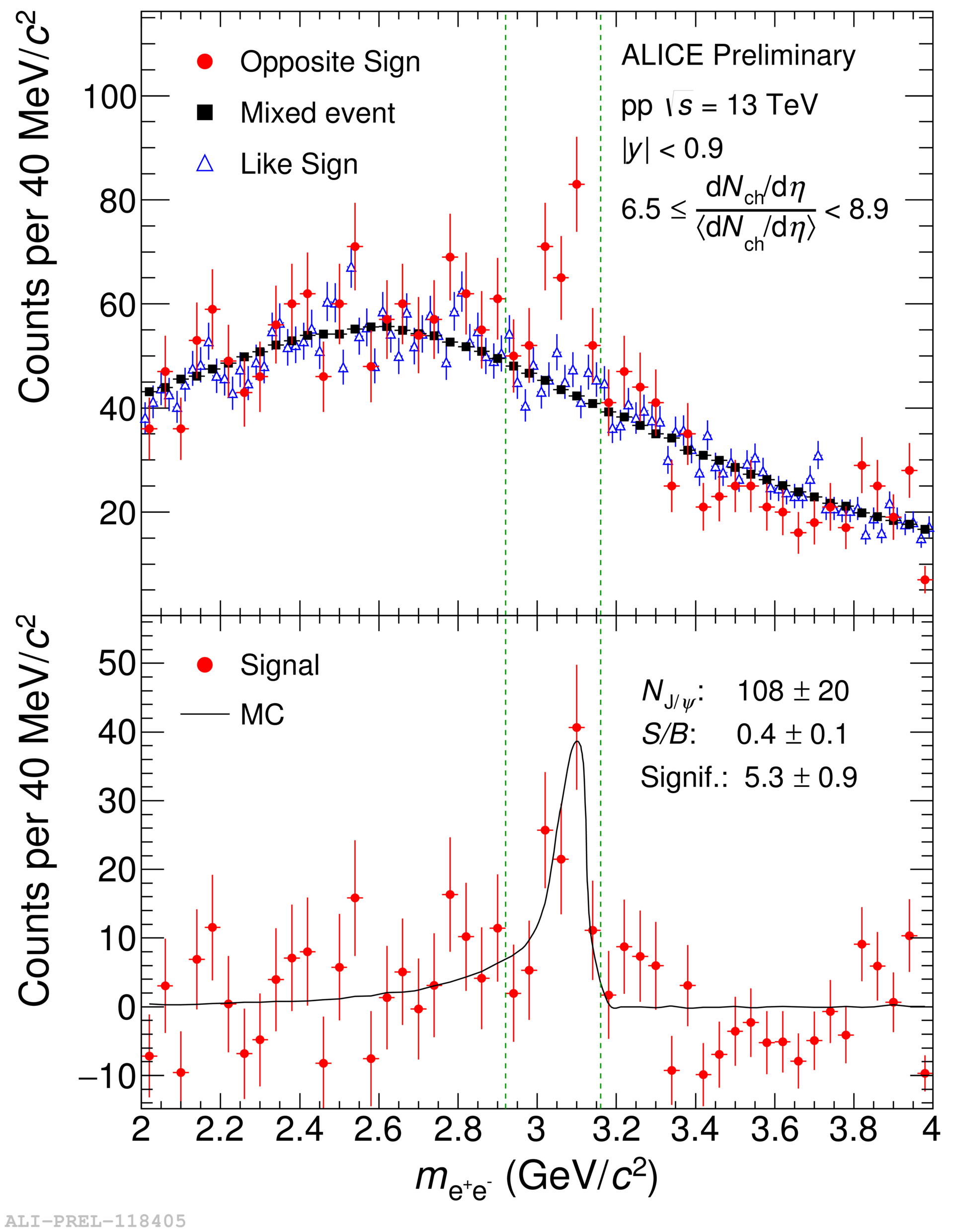}
\caption{Top: Invariant mass distribution of electron-positron pairs in the highest multiplicity bin, together with a background description using mixed-event or like-sign method. Bottom: Extracted signal; the black line indicates the expected line shape from simulations. \label{invmass}}
\end{minipage}
\end{figure}

The measurement was performed with the ALICE detector \cite{alice}. \jpsi particles were reconstructed at mid-rapidity ($|y| < 0.9$) in the dielectron decay channel. For tracking and selection of primary particles the Inner Tracking System (ITS) was used, and particle identification was done via specific energy loss in the Time Projection Chamber (TPC). Combinatorial background from electrons originating from photon conversions was suppressed by usage of a prefilter on dielectron pairs with a low opening angle.

In order to have access to high multiplicities, the data taking was performed using both a minimum bias trigger and a trigger on high event multiplicities, based on a large deposited charge in the ALICE V0 sub-detector. This detector consists of two scintillator arrays at forward ($2.8 < \eta < 5.1$) and backward ($-3.7 < \eta < -1.7$) rapidity. The analyzed data set corresponds to an integrated luminosity of about $\mathscr{L}=1.9\,\mathrm{nb}^{-1}$ for minimum bias events and $\mathscr{L}=1.2\,\mathrm{pb}^{-1}$ for high multiplicity triggered events. Pile-up events were rejected if a second primary vertex with sufficient contributors could be reconstructed within one event. From a toy Monte Carlo study it is estimated that the remaining pile-up contribution is negligible in all multiplicity bins.

The charged-particle multiplicity was estimated by the number of tracklets in the two innermost layers of the ITS. The distribution of the self-normalized charged-particle multiplicity for the events used in the analysis is shown in Fig. \ref{mult}, for minimum bias and high-multiplicity triggered events. Taking advantage of the triggering, the reach in multiplicity could be substantially extended compared to previous measurements at lower energies that used minimum bias events alone.

The signal was extracted from the invariant mass distribution by bin counting in the \jpsi signal region ($2.92 < m_{e^+e^-} < 3.16\,\textrm{GeV}/c^2$) after subtracting the combinatorial background, which was estimated with the event-mixing technique. As a cross-check also the distribution of like-sign pairs from the same event was used as a background estimator. The signal is extracted in 8 bins of the charged-particle multiplicity (indicated by the vertical lines in Fig. \ref{mult}), the highest one corresponding to about 7 times the mean values.

Figure \ref{invmass} shows the invariant mass distribution and the extracted signal in the bin corresponding to the highest multiplicity. In a final step the extracted yield and event multiplicities are normalized to the mean values in minimum bias events, obtained in the same analysis.

\section{Results}

\begin{figure}[t]
\begin{minipage}[b][][b]{0.5\textwidth}
\includegraphics[width=\textwidth]{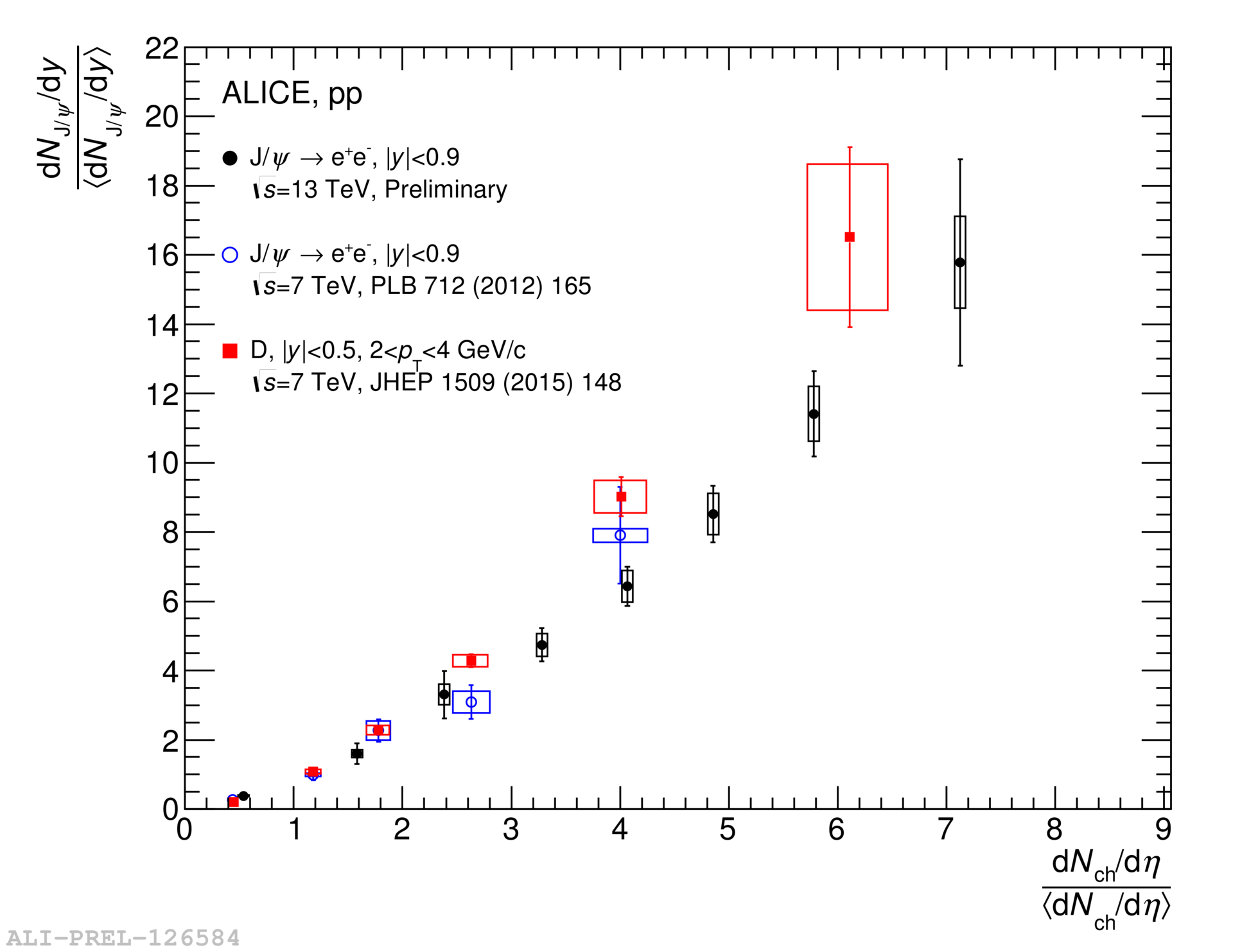}
\end{minipage}
\begin{minipage}[b][][b]{0.5\textwidth}
\includegraphics[width=\textwidth]{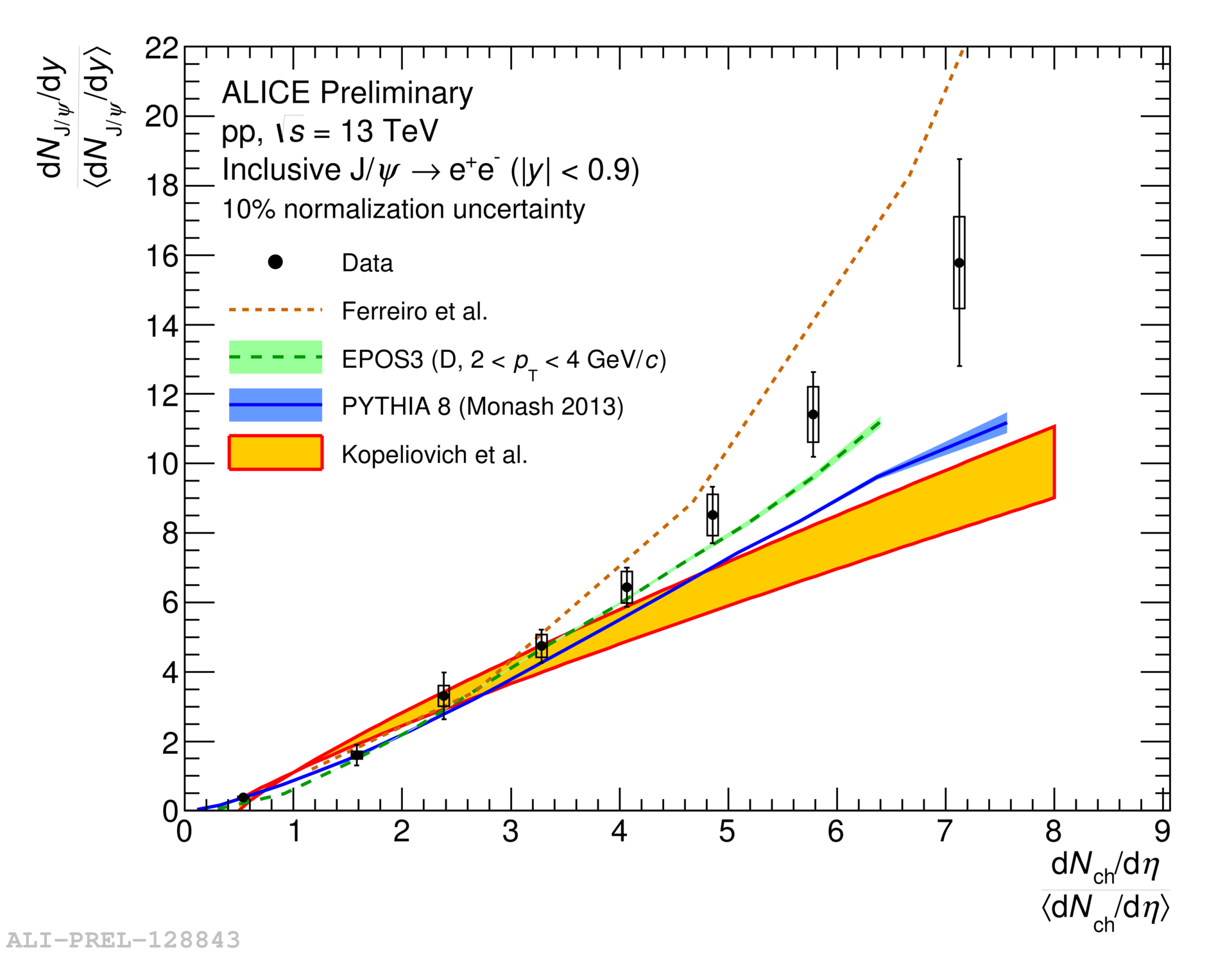}
\end{minipage}
\caption{Multiplicity dependence of inclusive \jpsi production at mid-rapidity at \sqrts{13}. Left: Comparison to results of D-meson and inclusive \jpsi production at mid-rapidity at $\sqrt{s}= 7$ TeV. Right: Comparison to predictions from PYTHIA 8 \cite{pythia}, EPOS 3 \cite{epos}, Ferreiro et al. \cite{percolation} and Kopeliovich et al. \cite{kopeliovich}.}
\label{result}
\end{figure}

Figure \ref{result} shows the self-normalized inclusive \jpsi yield (i.e. the yield per event divided by the mean yield per event in minimum bias collisions) in proton-proton collisions at \sqrts{13} at mid-rapidity as a function of the self-normalized charged-particle multiplicity. A smooth monotonic increase in the \jpsi yield is observed, which at high multiplicities is clearly stronger than the increase in the charged-particle multiplicity. The \jpsi yield reaches $15.4 \pm 2.9$ times the mean value at a multiplicity of 7.13 times the mean value.

In Fig. \ref{result} (left), the current results are compared to the measurements of \jpsi and D-meson production at \sqrts{7}. The \jpsi results are in mutual agreement, indicating that no significant energy dependence of the relative increase of \jpsi production with event multiplicity could be observed.

In Fig. \ref{result} (right), the results are compared to predictions from PYTHIA 8, EPOS 3, the percolation model, and the model based on higher Fock state contributions. EPOS3 and the percolation model predict a stronger than linear increase at high multiplicities. EPOS 3 provides a very good description of the data, the percolation model seems to slightly overestimate the increase at the highest multiplicities, PYTHIA8 and the model based on higher Fock states underestimates it.

\section{Conclusions}

ALICE has measured inclusive \jpsi production at mid-rapidity as a function of event multiplicity in pp collisions at \sqrts{13}. A monotonic increase with increasing event multiplicity has been observed. Comparing to the results obtained at \sqrts{7}, no energy dependence of the relative increase with multiplicity was observed.

The result is well described by the EPOS 3 event generators. This hints to an important role of multiple interactions on the parton level for charm production in hadron-hadron collisions.

At the highest multiplicities slight tensions exist between the data and the percolation model predictions, which overestimates the increase, as well as with PYTHIA8 predictions and the predictions from the model including higher Fock state contributions, which underestimate the increase. This might hint to a break-down of the model assumptions in this region or to the onset of other phenomena.





\bibliographystyle{elsarticle-num}
\bibliography{literatur}







\end{document}